\documentclass[hyper,letterpaper,notoc]{JHEP3}

\usepackage{epsfig}
\usepackage{amsbsy}
\usepackage{varioref}
\usepackage{pifont}
\usepackage{amsmath}

\title{Ideal Fermion Delocalization in Higgsless Models 
}

\author{R. Sekhar Chivukula and Elizabeth H. Simmons\\
Department of Physics and Astronomy, Michigan State University\\
East Lansing, MI 48824, USA\\
	E-mail: \email{sekhar@msu.edu, esimmons@msu.edu}}
	
\author{Hong-Jian He\\
Department of Physics, University of Texas\\
Austin, TX 78712, USA\\
	E-mail: \email{hjhe@physics.utexas.edu}}

\author{Masafumi Kurachi and Masaharu Tanabashi\\
Department of Physics, Tohoku University\\
Sendai 980-8578, Japan\\
	E-mail: \email{kurachi@tuhep.phys.tohoku.ac.jp, tanabash@tuhep.phys.tohoku.ac.jp}}

\abstract{
In this note we examine the properties of deconstructed Higgsless models
for the case of a fermion whose $SU(2)$ properties arise from delocalization over 
many sites of the deconstructed lattice.   We derive expressions for the correlation functions 
and use these to establish a generalized consistency relation among correlation functions.  We discuss 
the form of the $W$ boson wavefunction and show that if the probability distribution
of the delocalized fermions is appropriately related to the $W$ wavefunction, then deviations in precision electroweak parameters are minimized. In particular, we show that this ``ideal fermion
delocalization'' results in the vanishing of three of the four leading zero-momentum
electroweak parameters defined by Barbieri, et. al. We then discuss ideal fermion 
delocalization in the context of  two continuum Higgsless models, one in Anti-deSitter
space  and one in flat space. Our results may be applied to any Higgsless linear moose 
model with multiple $SU(2)$ groups, including those with only a few extra vector bosons.
}

\keywords{Dimensional Deconstruction, Electroweak Symmetry Breaking, Higgsless Theories, Delocalization}

\preprint{MSUHEP-050415\\
TU-742}

\begin{document}

\section{Introduction}

Higgsless models \cite{Csaki:2003dt} incorporate a mechanism for electroweak symmetry
breaking without a conventional scalar Higgs particle.   
The most popular models \cite{Agashe:2003zs,Csaki:2003zu} are based on a five-dimensional
$SU(2) \times SU(2) \times U(1)$ gauge theory in a slice of Anti-deSitter space, and
electroweak symmetry breaking is encoded in the boundary conditions of the
gauge fields on this space. The resulting spectrum includes the massless photon,
the $W$ and $Z$ bosons (which are the first Kaluza-Klein excitations of the five-dimensional
gauge fields), and an infinite tower of additional massive vector bosons (the remaining
``KK'' excitations). The unitarity of longitudinal $W$ and $Z$ boson scattering is ensured by the 
exchange of these other heavy vector bosons 
\cite{SekharChivukula:2001hz,Chivukula:2002ej,Chivukula:2003kq,He:2004zr}, 
rather than through the exchange
of a scalar Higgs boson \cite{Higgs:1964ia}. 
There have been numerous
studies of electroweak properties and collider phenolmenology
\cite{Cacciapaglia:2004jz,Nomura:2003du,Barbieri:2003pr,Davoudiasl:2003me,Burdman:2003ya,Davoudiasl:2004pw,Barbieri:2004qk,Hewett:2004dv} in the context of these five-dimensional models.

An alternative approach to analyzing the properties of Higgsless models
\cite{Foadi:2003xa,Hirn:2004ze,Casalbuoni:2004id,Chivukula:2004pk,Perelstein:2004sc,Chivukula:2004af,Georgi:2004iy,SekharChivukula:2004mu}
is to use deconstruction 
\cite{Arkani-Hamed:2001ca,Hill:2000mu} and to 
compute the electroweak parameters $\alpha S$ and $\alpha T$  
\cite{Peskin:1992sw,Altarelli:1990zd,Altarelli:1991fk} in a 
related linear moose model \cite{Georgi:1985hf}. We have recently shown 
\cite{SekharChivukula:2004mu} how to compute 
all four of the leading zero-momentum electroweak parameters defined
by Barbieri et. al. \cite{Barbieri:2004qk} in a very general class of linear moose models.
Using deconstruction, we are able to relate the size of these electroweak
corrections directly to the spectrum of the KK modes, which is constrained by
unitarity. Taking the continuum limit, our results apply directly to models with arbitrary
background 5-D geometry, spatially  dependent gauge-couplings, and
brane kinetic energy terms. 

Using the deconstruction approach, we first studied
Higgsless models with localized fermions,
{\it i.e.} fermions which derive their $SU(2)$ and $U(1)$ properties from
a single site on the deconstructed lattice.  We found \cite{SekharChivukula:2004mu} that any model of this kind which
does not have extra  light vector bosons (with masses of order the $W$ and $Z$) 
cannot simultaneously satisfy unitarity bounds and the constraints of precision electroweak data. Our analyses also apply directly to a large
class of models of extended electroweak symmetry which have only a few non-standard vector bosons
\cite{Chivukula:2003wj,Casalbuoni:1985kq,Casalbuoni:1996qt}; these models are motivated in
part by models of hidden local symmetry 
\cite{Bando:1985ej,Bando:1985rf,Bando:1988ym,Bando:1988br,Harada:2003jx}.

It has recently been proposed \cite{Cacciapaglia:2004rb,Foadi:2004ps}
 that the size of corrections to electroweak
processes may be reduced by including delocalized fermions. In deconstruction, a delocalized fermion is realized as a fermion whose $SU(2)$ properties arise from several
sites on the deconstructed lattice \cite{Chivukula:2005bn,Casalbuoni:2005rs}. 
We have previously considered
in detail the case of a fermions whose $SU(2)$ properties arise from
two adjacent sites \cite{Chivukula:2005bn}, and have confirmed that (even in that simple
case) it is possible to minimize the electroweak parameter $\alpha S$ by 
choosing a suitable amount of fermion delocalization. 

In this work we examine in detail the properties of deconstructed Higgsless models
for the case of a fermion whose $SU(2)$ properties arise from delocalization over 
many sites of the deconstructed lattice.   We derive explicit expressions for the correlation functions 
and use these to establish a generalized consistency relation among them.  We discuss 
the form of the $W$ boson wavefunction and show that if the probability distribution
of the delocalized fermions is appropriately related to the $W$ wavefunction, deviations in precision electroweak parameters are minimized. In particular, we show that this ``ideal fermion
delocalization'' results in the vanishing of three of the four leading zero-momentum
electroweak parameters defined by Barbieri, et. al. \cite{Barbieri:2004qk}. 
We then briefly discuss\footnote{
A detailed description of the deconstructed models corresponding in the continuum
limit to ideally delocalized fermions in flat and warped space will be deferred to
a subsequent work \cite{Chivukula:unpublished}.}  ideal fermion 
delocalization in the context of  two continuum Higgsless models, one in Anti-deSitter space 
and one in flat space. Our results may be applied to any Higgsless linear moose 
model with multiple $SU(2)$ groups, including those with only a few extra vector bosons.

\section{Review of the Model and Notation}

\EPSFIGURE[t]{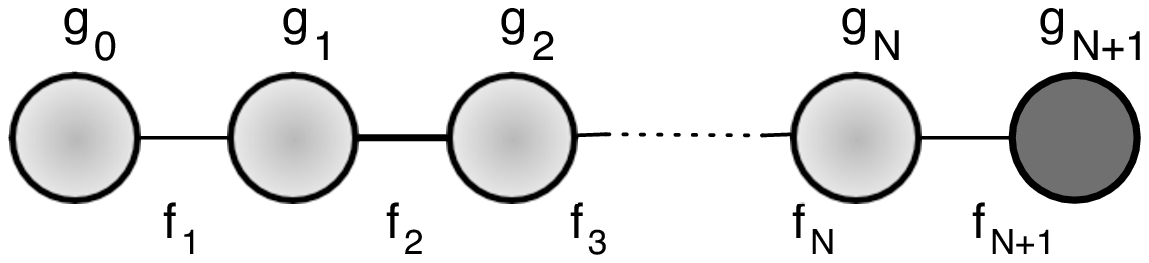,width=0.6\textwidth}
{Moose diagram of the model analyzed in this note. Sites $0$ to $N$ are
$SU(2)$ gauge groups, site $N+1$ is a $U(1)$ gauge group. The fermions are delocalized
in the sense that the $SU(2)$ couplings of the fermions
arise (potentially) from the gauge groups at all sites from 0 to $N$. The $U(1)$ coupling comes from the
gauge group at site $N+1$.
\label{fig:tone}}

We study a deconstructed Higgsless model,
as shown diagrammatically (using ``moose notation'' \cite{Georgi:1985hf}) in fig. \ref{fig:tone}.   The model incorporates an
$SU(2)^{N+1} \times U(1)$ gauge group, and $N+1$ 
nonlinear $(SU(2)\times SU(2))/SU(2)$ sigma models in which the global symmetry groups 
in adjacent sigma models are identified with the corresponding factors of the gauge group.
The Lagrangian for this model at leading order is given by
\begin{equation}
  {\cal L}_2 =
  \frac{1}{4} \sum_{j=1}^{N+1} f_j^2 \ \mbox{tr}\left(
    (D_\mu U_j)^\dagger (D^\mu U_j) \right)
  - \sum_{j=0}^{N+1} \dfrac{1}{2g_j^2} \ \mbox{tr}\left(
    F^j_{\mu\nu} F^{j\mu\nu}
    \right),
\label{lagrangian}
\end{equation}
with
\begin{equation}
  D_\mu U_j = \partial_\mu U_j - i A^{j-1}_\mu U_j 
                               + i U_j A^{j}_\mu,
\end{equation}
where all  gauge fields $A^j_\mu$ $(j=0,1,2,\cdots, N+1)$ are dynamical. The first
$N+1$ gauge fields ($j=0,1,\ldots, N$) correspond to $SU(2)$ gauge groups; the last gauge
field ($j= N+1$) corresponds to the  $U(1)$ gauge group.  The symmetry breaking between
the $A^{N}_\mu$ and $A^{N+1}_\mu$ follows an $SU(2)_L \times SU(2)_R/SU(2)_V$ symmetry
breaking pattern with the $U(1)$ embedded as the $T_3$-generator of $SU(2)_R$.   
Our analysis proceeds for arbitrary values of the gauge couplings and $f$-constants.
In the continuum limit, therefore, this allows for arbitrary background 5-D geometry,
spatially dependent gauge-couplings, and brane kinetic energy terms for the gauge-bosons. 

All four-fermion processes, including those relevant for the electroweak phenomenology of our model, depend, respectively, on the neutral and charged gauge field
propagator matrices
\begin{equation}
D^Z(Q^2) \equiv \left[ Q^2\, {\cal I} + M^2_{Z}\right]^{-1}~, \ \ \ \ \ \ \ \ 
{D}^W(Q^2) \equiv \left[ Q^2\, {\cal I} + {M}^2_{W}\right]^{-1}~.
\end{equation}
Here, $M_{Z}^2$ and ${M}_W^2$ are, respectively, the mass-squared matrices for the neutral and charged gauge bosons and ${\cal I}$ is the identity matrix.  Consistent with \cite{Chivukula:2004pk}, $Q^2 \equiv -q^2$ refers to the
Euclidean momentum. 

The neutral vector meson mass-squared matrix is of dimension $(N+2) \times (N+2)$ 
\begin{equation}
{\tiny
M_{Z}^2 = {1\over 4}
\left(
\begin{array}{c|c|c|c|c}
g^2_0 f^2_1& -g_0 g_1 f^2_1 & &  &  \\ \hline
-g_0 g_1 f^2_1  & g^2_1(f^2_1+f^2_2) & -g_1g_2 f^2_2&  &  \\ \hline
 & \ddots & \ddots & \ddots &  \\ \hline
 & & -g^{}_{N-1} g^{}_{N} f^2_{N} & g^2_{N}(f^2_{N} + f^2_{N+1}) & -g^{}_{N} g^{}_{N+1} f^2_{N+1}    \\ \hline
  & & & -g^{}_{N} g^{}_{N+1} f^2_{N+1} & g^2_{N+1}f^2_{N+1} 
\end{array}
\right).
}
\label{eq:neutralmatrix}
\end{equation}
and the charged current vector bosons' mass-squared matrix is the upper-left $(N+1)  \times (N+1) $ dimensional block of the  $M_{Z}^2$ matrix.
The neutral mass matrix (\ref{eq:neutralmatrix}) 
is of a familiar form that has a vanishing determinant, due to a zero eigenvalue.
Physically, this corresponds to a massless neutral gauge field -- the photon.
The non-zero eigenvalues of $M^2_{Z}$
are labeled by ${\mathsf m}^2_{Zz}$ ($z=0,1,2,\cdots, N$), while
those of ${M}^2_W$ are labeled by ${\mathsf m}^2_{Ww}$ ($w=0, 1,2,\cdots, N$). 

The lowest massive eigenstates corresponding to eigenvalues ${\mathsf m}^2_{Z0}$ and 
${\mathsf m}^2_{W0}$ are, respectively, identified as the usual $Z$ and $W$ bosons.
We will  refer to  these last eigenvalues by their conventional symbols $M^2_Z,\, M^2_{W}$; the distinction between these and the corresponding mass matrices should be clear from
context. 
We will denote the eigenvectors corresponding to the photon, $Z$, and $W$ by
$v^\gamma_i$, $v^Z_i$, and $v^W_j$. These eigenvectors are normalized as 
\begin{equation}
\sum_{i=0}^{N+1} (v^\gamma_i)^2 = \sum_{i=0}^{N+1} (v^Z_i)^2 = \sum_{j=0}^N (v^W_j)^2 = 1~.
\label{eq:evnorm}
\end{equation}
Inspection of the matrix $M^2_Z$ reveals that each component of the photon eigenvector is inversely related to the gauge coupling at the corresponding site
\begin{equation}
v^\gamma_i = {e\over g_i}~,\ \ \ \ {\rm where}\ \ \ \ \ \ {1\over e^2}=\sum_{i=0}^{N+1} {1\over g^2_i}~.
\label{eq:vgamma}
\end{equation}
In the continuum limit, the eigenstates with masses ${\mathsf m}^2_{Ww}$ and 
${\mathsf m}^2_{Zz}$ correspond to the higher Kaluza-Klein (``KK'') excitations of the
five-dimensional $W$ and $Z$ gauge fields.

Generalizing the usual mathematical
notation for ``open'' and ``closed'' intervals, we may denote \cite{SekharChivukula:2004mu} the
neutral-boson mass matrix $M^2_Z$ as $M^2_{[0,N+1]}$ --- {\it i.e.}
it is the mass matrix for the entire moose running from site $0$ to site $N+1$ including
the gauge couplings of both endpoint groups. Analogously, the charged-boson mass matrix $M^2_W$ is
$M^2_{[0,N+1)}$ --- it is the mass matrix for the moose running from site $0$ to link
$N+1$, but not including the gauge couping at site $N+1$.
This notation will be useful in thinking about the properties of sub-matrices $M^2_{[0,i)}$ of the full gauge-boson mass matrices that 
arise in our discussion of fermion delocalization, and also the corresponding eigenvalues 
${\mathsf m}^2_{i\,\hat{i}}\ (\hat{i}=1,2,\ldots, i)$.  We will denote the lightest such eigenvalue ${\mathsf m}^2_{i1}$ by the symbol $M^2_i$.

We will  find it useful to define the following sums over heavy eigenvalues for phenomenological discussions:
\begin{equation}
\Sigma_{Z} \equiv \sum_{z=1}^N  {1\over {\mathsf m}^2_{Zz}}\ ,\ \ \ \ \ 
\Sigma_{W} \equiv  \sum_{w=1}^N {1\over {\mathsf m}^2_{Ww}}\ ,\ \ \ \ \ 
\Sigma_{[0,i)}  \equiv {\rm Tr}\, M^{-2}_{[0,i)}~.
\label{eq:sig-def}
\end{equation}
That is, $\Sigma_Z$ and $\Sigma_W$ are the sums over inverse-square masses of the higher neutral- and charged-current KK modes of the full model.

\section{Deconstructed Delocalized Fermions}

The authors of \cite{Cacciapaglia:2004rb,Foadi:2004ps} consider the possibility that
the standard model fermions have wavefunctions with finite extent in the fifth dimension. 
In practice, this means
that the observed fermions are the lightest eigenstates of five-dimensional fermions, just as the $W$ and $Z$ gauge-bosons are the lightest in a tower of ``KK"
excitations. These authors show that by adjusting the five-dimensional wavefunction of the
light fermions, one can modify (and potentially eliminate) the dangerously large corrections
to precision electroweak measurements.  In this section, we establish what we mean by fermion delocalization in a deconstructed model; we explore the consequences in subsequent sections.

\subsection{Deconstructing fermion delocalization}
The deconstructed version of fermion delocalization proceeds very similarly to the continuum version.
Since a five-dimensional spinor is equivalent to a four-dimensional Dirac fermion, one introduces
a separate Dirac fermion at each site ({\it i.e.} one left-handed and one right-handed Weyl
spinor per site, $\psi^i_L$ and $\psi^i_R$) on the interior of the moose diagram of fig. \ref{fig:tone}.
The chirality of the standard model fermions is introduced by adjusting the boundary conditions for
the fermion fields at the ends of the moose. A convenient choice \cite{Cheng:2001nh} 
(consistent with the weak interactions) that we will adopt corresponds to 
\begin{equation}
\psi^{N+1}_L =0~,\qquad \psi^0_R = 0~.
\end{equation}
Discretizing the Dirac action for a five-dimensional fermion in an arbitrary background
metric then corresponds to introducing site-dependent masses ($m_j$) for the Dirac fermions at
each interior site and postition-dependent Yukawa 
interactions ($y_j$) which couple the left-handed modes at site 
$j$ to the right-handed modes at site $j+1$
\begin{equation}
{\cal L}_{5f} = - \sum_{j=1}^{N-1} m_j \bar{\psi}^j_L \psi^j_R 
- \sum_{j=0}^{N-1} f_{j+1}\, y_{j+1}\left( \overline{\psi^j_L} U_{j+1} \psi^{j+1}_R\right) + {h.c.}~,
\label{eq:fmatrix}
\end{equation}
where gauge-invariance dictates that each such interaction include a factor 
of the link field $U_{j+1}$, and we therefore write the corresponding interaction proportional
to $f_{j+1}$.  

Note that in eqn. (\ref{eq:fmatrix}) we have not included a Yukawa coupling corresponding to
link $N+1$. Given that there is only a $U(1)$ interaction at site $N+1$, there are actually two
possible such link terms (corresponding to the two Yukawa couplings of up- and down-type
fermions in the standard model). We will analyze the model in the limit where the lightest fermion eigenstates (which we identify with the standard model fermions) are massless.  
The absence of the  Yukawa couplings at site $N+1$  insures\footnote{This corresponds to
the limit in which the coupling $t_R \to 0$ in ref. \protect\cite{Foadi:2004ps}.} 
that the right-handed components of these massless modes are localized 
entirely at site $N+1$. For simplicity, in what follows we will also assume flavor universality, 
{\it i.e.} that the same five-dimensional fermion mass matrix applies to all flavors
of fermions.

In this limit, only the left-handed components of the massless
fermions are delocalized, and their behavior is characterized by a wavefuntion
\begin{equation}
|\psi_L \rangle = \left(
\begin{array}{c}
\alpha_0 \\
\alpha_1 \\
\vdots \\
\alpha_N
\end{array}
\right)~,
\end{equation}
where the $\alpha_j$ are complex parameters.  Denoting $\vert \alpha_i\vert^2 \equiv x_i$ and recognizing that 
\begin{equation}
\sum_{i=0}^N x_i = 1~.
\label{eq:xsum}
\end{equation}
we find that the couplings of the ordinary (zero-mode) fermions
in this model may be written
\begin{equation}
{\cal L}_f = \vec{J}^\mu_L \cdot \left( \sum_{i=0}^N x_i \vec{A}^i_\mu \right)
+ J^\mu_Y A^{N+1}_\mu~.
\label{eq:fcoupling}
\end{equation}
As usual, $\vec{J}^\mu_L$ denotes the isotriplet of left-handed 
weak fermion currents and $J^\mu_Y$ is the fermion hypercharge current.

The values $x_i$ depend on the details of the model chosen -- in particular, on the form
of the bulk mass, warping, and boundary conditions chosen for the fermions \cite{Cacciapaglia:2004rb,Foadi:2004ps}.  For the first part of our discussion, we will consider the parameters $x_i$ as arbitrary.  Later on, we will focus on a particular choice for the $x_i$, a so-called ``ideal delocalization" which minimizes  the values of corrections to precision electroweak variables.

\subsection{Re-interpreting fermion delocalization}

As noted earlier in \cite{Chivukula:2005bn,Casalbuoni:2005rs}, the ``delocalized'' fermion coupling in deconstructed Higgsless models, 
eqn. (\ref{eq:fcoupling}), may also be written using the Goldstone
boson fields of the Moose in fig. 1. For each $j$, the non-linear sigma model
field $U_j$ in eqn. (\ref{lagrangian}) corresponds to link $j$ of the moose
and transforms under the adjacent $SU(2)_{j-1} \times SU(2)_j$
groups as   $U_j \to V_{j-1} U_j V_{j}^\dagger$.
Using these fields, define 
\begin{equation}
W_k = U_1\cdot U_2 \cdots U_k~,
\end{equation}
which  transforms as $W_k \to V_0 W_i V_{k}^\dagger$ under $SU(2)_0
\times SU(2)_k$. Consider the current operator
\begin{equation}
{\rm Tr} \left( {\sigma^a\over 2} W^\dagger_k iD_\mu W_k \right)\  \rightarrow\ 
+{1\over 2}(A^a_{0\mu} - A^a_{k\mu})~,
\end{equation}
where the $\sigma^a$ are the Pauli matrices, $D_\mu$ is the appropriate covariant derivative,
and where  we have specified the form of this operator in unitary gauge in which all the link
fields $U_j \equiv {\cal I}$.

In this language, the fermions' weak couplings may be written (using eqn. (\ref{eq:xsum}))
\begin{equation}
\vec{J}^\mu_L \cdot \left[ \vec{A}^0_\mu - \sum_{k=1}^N 2 x_k 
{\rm Tr} \left( {\vec{\sigma}\over 2} W^\dagger_k iD_\mu W_k \right) \right]~.
\label{eq:newcoupling}
\end{equation}
From this point of view, the fermions are charged only under $SU(2)_0$ and
the apparent delocalization comes about from couplings to the Goldstone-boson fields.

Note that, in the gauge-boson normalization we are using, the linear combinations of
gauge fields $A^a_{0\mu} - A^a_{k\mu}$ are strictly orthogonal to the photon
\begin{equation}
A^\gamma_{\mu} \propto A^3_{0\mu} + A^3_{1\mu} +\ldots + A^3_{N+1\, \mu}~.
\end{equation}
Hence, the couplings of eqn. (\ref{eq:newcoupling}) result in a modification of the
$Z$ and $W$-couplings whose size depends on the $x_k$ and the admixture of $A_0-A_k$
in the mass-eigenstate $W$ and $Z$ fields.  The couplings of eqn. (\ref{eq:newcoupling}) do not modify the photon coupling.

\section{Correlation Functions and Consistency Relations}

\subsection{Correlation Functions}

Recalling that fermions may be charged under any of the single $SU(2)$ gauge groups, as well as under the single $U(1)$ group at the $N+1$ site, neutral current four-fermion processes may be derived from the Lagrangian
\begin{eqnarray}
{\cal L}_{nc}  & = & - {1\over 2} \left[ \sum_{i,j = 0}^N x_i x_j g_i g_j \, D^Z_{i,j}(Q^2) \right] 
J^\mu_3 J_{3\mu} - \left[ \sum_{i=0}^N x_i g_i g_{N+1}\, D^Z_{i,N+1}(Q^2)\right]  J^\mu_3 J_{Y\mu} \nonumber \\
&  &  \quad - {1\over 2} \left[ g^2_{N+1}\, D^Z_{N+1,N+1}(Q^2)\right]  J^\mu_Y J_{Y\mu}~,
\label{nclagrangian}
\end{eqnarray}
and charged-current process from
\begin{equation}
{\cal L}_{cc} = - {1\over 2} \left[ \sum_{i,j=0}^{N} x_i x_j g_i g_j \, {D}^W_{i,j}(Q^2)\right]
 J^\mu_+ J_{-\mu}~.
\label{cclagrangian}
\end{equation}
where $D_{i,j}$ is the $(i,j)$ element of the appropriate gauge field propagator matrix.
We can define correlation functions between fermion currents at given sites as
\begin{equation}
[G_{NC}(Q^2)]_{i,j} = g_i g_j D^Z_{i,j} (Q^2)~,\qquad \qquad [G_{CC}(Q^2)]_{i,j} = g_i g_j D^W_{i,j}(Q^2)\ .
\end{equation}

The hypercharge correlation function $[G_{NC}(Q^2)]_{YY} = [G_{NC}(Q^2)]_{N+1,N+1}$ depends only on the single site with a $U(1)$ gauge group.   
This correlation function is the same as for the simplest ``Case I''
model with localized fermions discussed in \cite{SekharChivukula:2004mu} \footnote{A Case I model is a linear moose with a set of $SU(2)$ groups adjacent to a set of $U(1)$ groups; all fermions get their hypercharge from the $U(1)$ adjacent to the $SU(2)$ groups.}
\begin{equation}
[G_{\rm NC}(Q^2)]_{N+1,N+1}=[G_{\rm NC}(Q^2)]_{YY} = {e^2 M^2_Z (Q^2+M^2_W) \over Q^2 M^2_W (Q^2 + M^2_Z)}
\left[\prod_{w=1}^N {Q^2 + {\mathsf m}^2_{Ww} \over {\mathsf m}^2_{Ww}}\right]
\left[ \prod_{z=1}^N {{\mathsf m}^2_{Zz} \over Q^2 + {\mathsf m}^2_{Zz}}\right]
~.
\label{eq:gncyyi}
\end{equation}
The delocalization of the fermions has no effect on this correlation function.

The full correlation function for the fermion currents $J^\mu_3$ and $J^\mu_Y$ is 
\begin{equation}
[G_{\rm NC}(Q^2)]_{WY}= \sum_{i=0}^N x_i  [G_{\rm NC}(Q^2)]_{i,N+1} ~,
\label{eq:gncwyii}
\end{equation}
where we have used eqn. (\ref{eq:fcoupling}) to include the appropriate contribution from each site to which fermions couple.  By direct evaluation, following the analysis
of  \cite{SekharChivukula:2004mu}, 
we find the relevant elements of the propagator matrix have the form
\begin{eqnarray}
[G_{\rm NC}(Q^2)]_{0,N+1} & = & {e^2 M^2_Z \over Q^2 (Q^2 + M^2_Z)}
\left[ \prod_{z=1}^N {{\mathsf m}^2_{Zz} \over Q^2 + {\mathsf m}^2_{Zz}}\right] 
\label{eq:gnc0n}
\cr
[G_{\rm NC}(Q^2)]_{i,N+1} & = & {e^2 M^2_Z \over Q^2 (Q^2 + M^2_Z)}
\left[ \prod_{\hat{i}=1}^i \left(1+ {Q^2\over {\mathsf m}_{i\,\hat{i}}^2}\right)\right]
\left[ \prod_{z=1}^N {{\mathsf m}^2_{Zz} \over Q^2 + {\mathsf m}^2_{Zz}}\right]~.
\label{eq:gncin}
\end{eqnarray}
Finally, the full correlation functions for weak currents are
\begin{equation}
[G_{\rm NC, CC}]_{WW} = \sum_{i,j=0}^N x_i x_j [G_{\rm NC, CC}]_{i,j}~.
\label{eq:usefulsum}
\end{equation}
We discuss these correlation functions in subsection 4.3.

\subsection{Spectral Decomposition and Residue Consistency Relations}

Each correlation function may be written in a spectral decomposition in terms of the mass eigenstates and their corresponding pole residues, $\xi$, as follows:
\begin{equation}
[G_{\rm NC}(Q^2)]_{YY} =  \dfrac{[\xi_\gamma]_{YY}}{Q^2} 
+\dfrac{[\xi_Z]_{YY}}{Q^2 + M_Z^2} 
+\sum_{z=1}^{N} \dfrac{[\xi_{Zz}]_{YY}}{Q^2 + {\mathsf m}_{Zz}^2},
\label{eq:NC_YY} 
\end{equation}
\begin{equation} 
  [G_{\rm NC}(Q^2)]_{WY} = \dfrac{[\xi_\gamma]_{WY}}{Q^2} 
  +\dfrac{[\xi_Z]_{WY}}{Q^2 + M_Z^2} 
  +\sum_{z=1}^{N} \dfrac{[\xi_{Zz}]_{WY}}{Q^2 + {\mathsf m}_{Zz}^2},
\label{eq:NC_WY}  
\end{equation}
\begin{equation}
  [G_{\rm NC}(Q^2)]_{WW} =  \dfrac{[\xi_\gamma]_{WW}}{Q^2} 
 +\dfrac{[\xi_Z]_{WW}}{Q^2 + M_Z^2} 
  +\sum_{z=1}^{N} \dfrac{[\xi_{Zz}]_{WW}}{Q^2 + {\mathsf m}_{Zz}^2},
\label{eq:NC_WW}  
\end{equation}
\begin{equation}
  [G_{\rm CC}(Q^2)]_{WW} =  \dfrac{[\xi_W]_{WW}}{Q^2 + M_W^2} 
  +\sum_{w=1}^{N} \dfrac{[\xi_{Ww}]_{WW}} {Q^2 + {\mathsf m}_{Ww}^2},
\label{eq:CC_WW}  
\end{equation}
All poles should be simple (i.e. there should be no degenerate mass eigenvalues) because
we are analyzing the discrete version of a self-adjoint operator on a finite interval.  
Since the neutral bosons couple to only two physically distinct currents, $J^\mu_3$ and $J^\mu_Y$, 
the three sets of residues in equations (\ref{eq:NC_YY})--(\ref{eq:NC_WY}) must be related. 
Specifically, they satisfy the $N+1$ consistency conditions,
\begin{equation}
  [\xi_Z]_{WW} [\xi_Z]_{YY}
  = \left([\xi_Z]_{WY}\right)^2, \qquad
  [\xi_{Z{z}}]_{WW} [\xi_{Z{z}}]_{YY}
  = \left([\xi_{Z{z}}]_{WY}\right)^2 .
\label{consistency}
\end{equation}
In the case of the photon, charge universality further implies
\begin{equation}
  e^2 = [\xi_\gamma]_{WW} = [\xi_\gamma]_{WY} = [\xi_\gamma]_{YY}.
\label{eq:universality}
\end{equation}

Finally, we note that the residues of the poles appearing in the spectral representations
of eqns. (\ref{eq:NC_YY}) -- (\ref{eq:CC_WW}) are directly related to the gauge boson eigenvectors,
$v^W_i$ and $v^Z_i$.   To see this, recall that each residue is the product of the couplings
of the related gauge boson to the appropriate fermion currents
\begin{align}
[\xi_W]_{WW} & = g^2_W~, & [\xi_Z]_{WW} & = (g^W_Z)^2~,\label{eq:xivs}\\
[\xi_Z]_{WY} & =g^W_Z g^Y_Z~, & [\xi_Z]_{YY} & = (g^Y_Z)^2~.
\end{align}
(which also leads back to the consistency relations in (\ref{consistency}) above).  
The coupling of a gauge boson mass eigenstate to a fermion current is the sum of the contributions from each site 
\begin{align}
g_W & = \sum_{j=0}^N x_j g_j v^W_j~,\\
g^W_Z & = \sum_{j=0}^N x_j g_j v^Z_j~,\\
g^Y_Z & = g_{N+1} v^Z_{N+1}~,
\end{align}
and therefore reflects the fermion and gauge boson wave-functions and the site-dependent couplings.
So the residues are indeed related to the eigenvectors.

\subsection{Consistency Relations Among Correlation Functions}

\EPSFIGURE[ht]{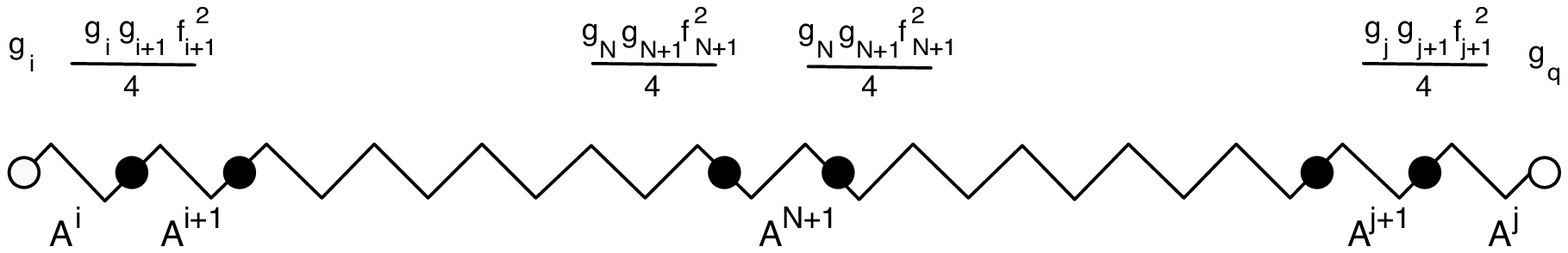,width=0.9\textwidth}
{Leading diagram at high-$Q^2$ ($Q^2\gg {\mathsf m}^2_{Zz,Ww}$
for $w,z>0$) which 
distinguishes $[G_{\rm NC}(Q^2)]_{i,j}$ from
$[G_{\rm CC}(Q^2)]_{i,j}$.
\label{fig:ttwo}
}

Consider the difference of charged and neutral current correlation functions
\begin{equation}
[G_{\rm NC}(Q^2)]_{i,j} - [G_{\rm CC}(Q^2)]_{i,j}~,
\end{equation}
where, without loss of generality, we may take $i \le j$.
At high-$Q^2$, the leading contribution to the difference is shown in
fig. \ref{fig:ttwo}, and therefore at large momenta
\begin{equation}
[G_{\rm NC}(Q^2)]_{i,j} - [G_{\rm CC}(Q^2)]_{i,j} \propto {g^2_{N+1} \over (Q^2)^{2N-i-j+3}}~.
\end{equation}
The only consistent way to achieve the high energy behavior 
while including the poles appropriate for the charged and 
neutral correlation functions is to write 
\begin{equation}
  [G_{\rm NC}(Q^2)]_{i,j}  - [G_{\rm CC}(Q^2)]_{i,j} 
= \dfrac{e^2 M_W^2 M_Z^2 \,{\cal R}(Q^2)}{Q^2[Q^2+M_W^2][Q^2 + M_Z^2]}
        \left[
        \prod_{{w}=1}^{N} 
        \dfrac{{\mathsf m}_{W{w}}^2}
              {Q^2 + {\mathsf m}_{W{w}}^2}
        \right]
         \left[
        \prod_{{z}=1}^{N} 
        \dfrac{{\mathsf m}_{Z {z}}^2}
              {Q^2 + {\mathsf m}_{Z{z}}^2}
        \right]~,
\label{eq:exact_WW1}
\end{equation}
where the function ${\cal R}(Q^2)$ is a polynomial in $Q^2$ of order $i+j$ and 
${\cal  R}(0) = 1$ in order to satisfy charge universality. Furthermore, by direct evaluation
 \cite{SekharChivukula:2004mu}, we see that
\begin{equation}
[G_{\rm NC,CC}(Q^2)]_{i,j} \propto \left[ \prod_{\hat{i}=1}^i \left(1+ {Q^2\over {\mathsf m}_{i\,\hat{i}}^2}\right)\right] ~,
\end{equation}
and therefore the polynomial satisfies the $i$ conditions ${\cal R}(-{\mathsf m}^2_{i\,\hat{i}}) = 0$.

Recall  the correlation functions $[G_{\rm NC}(Q^2)]_{i,N+1}$ and $[G_{\rm NC}(Q^2)]_{j,N+1}$,
which can be written as in eqn. (\ref{eq:gncin}), and the correlation function
$[G_{\rm NC}(Q^2)]_{N+1,N+1} \equiv [G_{\rm NC}(Q^2)]_{YY}$.   We may expand each of these, as well as $[G_{\rm NC}(Q^2)]{i,j}$ in a spectral representation like the one employed in eqns. (\ref{eq:NC_YY} - \ref{eq:CC_WW}) .  The residues of these correlation functions will therefore satisfy the $N+1$ conditions
\begin{equation}
[\xi_Z]_{i,j} ={[\xi_Z]_{i,N+1} [\xi_Z]_{j,N+1} \over [\xi_Z]_{N+1,N+1}}~, \quad
[\xi_{Zz}]_{i,j} ={[\xi_{Zz}]_{i,N+1} [\xi_{Zz}]_{j,N+1} \over [\xi_{Zz}]_{N+1,N+1}}~,
\end{equation}
The combination of these $N+1$ conditions and the $i$ 
conditions ${\cal R}(-{\mathsf m}^2_{i\,\hat{i}}) = 0$ overdetermines
${\cal R}(Q^2)$, and we find a consistency condition among correlation functions\footnote{This
expression can also be derived from the consistency relation given in 
\protect\cite{SekharChivukula:2004mu} for  $[G_{\rm NC}(Q^2)]_{j,j} - [G_{\rm CC}(Q^2)]_{j,j}$, 
and using the explicit forms of the correlation functions.}

\begin{equation}
[G_{\rm NC}(Q^2)]_{i,j}  - [G_{\rm CC}(Q^2)]_{i,j} =
{[G_{\rm NC}(Q^2)]_{i,N+1} [G_{\rm NC}(Q^2)]_{j,N+1} \over
[G_{\rm NC}(Q^2)]_{N+1,N+1}}~.
\label{eq:consistency}
\end{equation}
Direct evaluation of the correlation functions on the right hand side of this expression
(using eqns. (\ref{eq:gncyyi}) and (\ref{eq:gncin})) 
yields\footnote{Here ${\mathsf m}^2_{j\,\hat{j}}$ are the eigenvalues of
$M^2_{[0,j)}$.}
\begin{eqnarray}
\lefteqn{
  [G_{\rm NC}(Q^2)]_{i,j} 
 -[G_{\rm CC}(Q^2)]_{i,j}  = \dfrac{e^2 M_Z^2 M_W^2}{Q^2(Q^2+M_Z^2)(Q^2+M_W^2)} \times
} \nonumber\\
 & & 
     \left[
       \prod_{z=1}^{N}
       \dfrac{{\mathsf m}^2_{Z_{z}}}{Q^2 + {\mathsf m}^2_{Z_{z}}}
     \right]
     \left[
       \prod_{w=1}^{N}
       \dfrac{{\mathsf m}^2_{W_{w}}}{Q^2 + {\mathsf m}^2_{W_{w}}}
     \right]    \left[ \prod_{\hat{i}=1}^i \left(1+ {Q^2\over {\mathsf m}_{i\,\hat{i}}^2}\right)\right]
    \left[ \prod_{\hat{j}=1}^j \left(1+ {Q^2\over {\mathsf m}_{j\,\hat{j}}^2}\right)\right]
    \label{eq:consistent}
     \end{eqnarray}

Finally,  applying eqn. (\ref{eq:usefulsum}) to the LHS of (\ref{eq:consistent}) and eqns. (\ref{eq:gncwyii}) and (\ref{eq:gncyyi}) to its RHS reveals that the full correlation functions satisfy the same consistency relations %
\begin{equation}
[G_{\rm NC}(Q^2)]_{WW}  - [G_{\rm CC}(Q^2)]_{WW} =
{([G_{\rm NC}(Q^2)]_{WY})^2 \over
[G_{\rm NC}(Q^2)]_{YY}}
\label{eq:consistencyi}
\end{equation}
as those \cite{SekharChivukula:2004mu}
 in Case I models with fermions localized at a single site.

\subsection{$W$ Eigenvector}
\label{sec:weigenvector}

Since the residues of the poles of the correlation functions are related to the
gauge couplings ($g_i$) and eigenvectors ($v_i$), as in Section 4.2, we may use our expressions 
for the correlation functions to find the components of the massive eigenvectors 
in terms of the spectrum and gauge couplings. Here, we derive the
form of the $W$ eigenvector from the consistency condition of eqn. (\ref{eq:consistent}) and the relation
\begin{equation}
[\xi_W]_{i,j}  =  g_i g_j v^W_i v^W_j~,
\label{eq:xiggv}
\end{equation}
Computing the residue of the $W$ pole in (\ref{eq:consistent}) and expanding it in inverse powers of 
the higher KK masses
we find

\begin{equation}
[\xi_W]_{i,j}  =   \left({e^2 \over {1-{M^2_W\over M^2_Z}}}\right)
\left[1+M^2_W (\Sigma_W+\Sigma_Z)\right]
\left[ \prod_{\hat{i}=1}^i \left(1-{M_W^2\over {\mathsf m}_{i\,\hat{i}}^2}\right)\right]
\left[ \prod_{\hat{j}=1}^j \left(1-{M_W^2\over {\mathsf m}_{j\,\hat{j}}^2}\right)\right]~,
\end{equation}
and hence, using (\ref{eq:xiggv}), 
\begin{equation}
g_i v^W_i = {e\over \sqrt{1-{M^2_W\over M^2_Z}}}
\left[1+{M^2_W \over 2} (\Sigma_W+\Sigma_Z)\right]
\left[ \prod_{\hat{i}=1}^i \left(1-{M_W^2\over {\mathsf m}_{i\,\hat{i}}^2}\right)\right]~.
\end{equation}
Note that $v^W_{N+1} \equiv 0$, as it must be given the form of the model in
fig. \ref{fig:tone}.  Maintaining positive values for the remaining elements of the $v^W$ eigenvector (corresponding to a nodeless wavefunction in the continuum limit) implies
that $M^2_W < {\mathsf m}^2_{i\,\hat{i}}$ for all $i$ and $\hat{i}$.

For $i$ large, close to $N+1$, we expect that the matrix $M^2_{[0,i)}$ will
have a light eigenvalue close to $M^2_W$. For $i$ small, however, we expect
that we may expand the product in powers of $M^2_W / {\mathsf m}_{i\,\hat{i}}^2$ and find the approximate form
\begin{equation}
g_i v^W_i = {e\over \sqrt{1-{M^2_W\over M^2_Z}}}
\left[1+{M^2_W \over 2} (\Sigma_W+\Sigma_Z)-M_W^2\Sigma_{[0,i)}\right]~.
\end{equation}
For small $i$ therefore,  to leading order in the inverse mass-squared expansion,
$g_i v^W_i $ is approximately a constant.

\section{Ideal Delocalization}

A big question remains: what kinds of delocalization schemes produce viable models? Viability depends, in part, on satisfying constraints from precision electroweak corrections, so we remind the reader of the relevant definitions.  Then we introduce ``ideal delocalization" which guarantees that
the corrections can be made small.  We discuss this in deconstructed language.  Then in section 6, we show that the results hold neatly in the continuum.

\subsection{Electroweak Parameters}

As we have shown in \cite{Chivukula:2004af}, the most
 general amplitude for low-energy four-fermion neutral weak current processes in
any ``universal'' model \cite{Barbieri:2004qk} may be written 
as\footnote{See \cite{Chivukula:2004af} for a discussion of the correspondence 
between the ``on-shell'' parameters defined here, and the zero-momentum
parameters defined in  \protect{\cite{Barbieri:2004qk}}.  Note that $U$ is shown in \cite{Chivukula:2004af} to be zero to the order we consider in this paper.}
\begin{eqnarray}
-{\cal M}_{NC} = e^2 {{\cal Q}{\cal Q}' \over Q^2} 
& + &
\dfrac{(I_3-s^2 {\cal Q}) (I'_3 - s^2 {\cal Q}')}
	{\left({s^2c^2 \over e^2}-{S\over 16\pi}\right)Q^2 +
		{1\over 4 \sqrt{2} G_F}\left(1-\alpha T +{\alpha \delta \over 4 s^2 c^2}\right)
		} 
\label{eq:NC4} \\ \nonumber & \ \ & \\
&+&
\sqrt{2} G_F \,{\alpha \delta\over  s^2 c^2}\, I_3 I'_3 
+ 4 \sqrt{2} G_F  \left( \Delta \rho - \alpha T\right)({\cal Q}-I_3)({\cal Q}'-I_3')~,
\nonumber 
\end{eqnarray}
and the matrix element for charged current process may be written 
\begin{eqnarray}
  - {\cal M}_{\rm CC}
  =  \dfrac{(I_{+} I'_{-} + I_{-} I'_{+})/2}
             {\left(\dfrac{s^2}{e^2}-\dfrac{S}{16\pi}\right)Q^2
             +{1\over 4 \sqrt{2} G_F}\left(1+{\alpha \delta \over 4 s^2 c^2}\right)
            }
        + \sqrt{2} G_F\, {\alpha  \delta\over s^2 c^2} \, {(I_{+} I'_{-} + I_{-} I'_{+}) \over 2}~.
\label{eq:CC3}
\end{eqnarray}
The parameter $s^2$ is defined implicitly in these expressions as the ratio of
the ${\cal Q}$ and $I_3$ couplings of the $Z$ boson.
$\Delta \rho$ corresponds to the deviation from unity of the ratio of the strengths of
low-energy isotriplet weak neutral-current scattering and charged-current scattering.
$S$ and $T$ are the familiar oblique electroweak parameters \cite{Peskin:1992sw,Altarelli:1990zd,Altarelli:1991fk}, 
as determined by examining the {\it on-shell} properties of the $Z$ and $W$ bosons.
The contact interactions proportional to $\alpha \delta$ and ($\Delta \rho - \alpha T$)
correspond to ``universal non-oblique'' corrections arising from the exchange of
heavy KK modes. 
Finally, the consistency relation, eqn. (\ref{eq:consistencyi}), insures
that $\Delta \rho=0$ in any Case I model
\cite{Chivukula:2004pk}, regardless of fermion
delocalization.

\subsection{Optimizing the Fermion Wavefunction: $\alpha \delta$}

We now consider how to choose the form of the delocalized fermion
wavefunction so as to minimize the deviations in the electroweak parameters. Consider
first the parameter $\alpha \delta$ --  from the form of eqn. (\ref{eq:CC3})
we see that this deviation arises from the exchange of massive KK modes.
Therefore,
\begin{equation}
  \alpha\delta \propto \sum_{w=1}^N  \sum_{i,j} x_i x_j 
     \dfrac{[\xi_{W_{w}}]_{i,j}}{{\mathsf m}^2_{W_{w}}},
\end{equation}
where $[\xi_{W_{w}}]_{i,j}$ is the pole residue of 
$[G_{\rm CC}(Q^2)]_{i,j}$ at $Q^2 = {\mathsf m}^2_{W_{w}}$.
In analogy to eqn. (\ref{eq:xiggv}), the pole residues are 
related to the $W_{w}$ wave functions $v_i^{W_{w}}$, 
\begin{equation}
  [\xi_{W_{w}}]_{i,j} = g_i g_j 
      v_i^{W_{w}} v_j^{W_{w}}.
\end{equation}
For a given fermion wavefunction $x_i$, we thus find
\begin{equation}
  \alpha\delta \propto 
  \sum_{w=1}^N
  \dfrac{\left(\sum_{i} g_i x_i v_i^{W_{w}}\right)^2}
        {{\mathsf m}^2_{W_{w}}}~,
\end{equation}
and $\alpha\delta$ is therefore a positive-semi-definite parameter.

To minimize $\alpha\delta$, we exploit the fact that the eigenvector for
the massive $W$ and those for each of the KK modes are mutually orthogonal
\begin{equation}
  \sum_i v_i^{W} v_i^{W_{w}} = 0~.
\end{equation}
Since $\alpha\delta$ involves the product $x_i v_i^{W_{w}}$, we 
choose our ``ideally delocalized" fermion wavefunction $x_i$ to be
related to the form of the $W$ wavefunction 
\begin{equation}
  g_i x_i = {\cal N}  v_i^{W}~,
\label{eq:ideal}
\end{equation}
where the normalization factor ${\cal N}$ is fixed by the constraint $\sum_i x_i=1$.
By construction, the contact interaction $\alpha\delta$ vanishes for
an ideally delocalized fermion.
As noted earlier, we may choose the lightest gauge boson eigenvector to be positive definite (corresponding to a nodeless
wavefunction in the continuum limit), and therefore 
$x_i \ge 0$ for ideal delocalization.

\subsection{$\alpha S$ and $\alpha T$ for an Ideally Delocalized Fermion}

Consider the following combination of correlation functions with the
fermion wavefunction
\begin{equation}
  \sum_{i} x_i \left(
    [G_{\rm NC}(Q^2)]_{i,j} 
   -[G_{\rm CC}(Q^2)]_{i,j}
  \right).
\label{eq:diff1}
\end{equation}
If we assume an ideally delocalized fermion, as defined above, it cannot couple to the higher 
charged KK modes. Therefore the combination of correlation functions  in  eqn. (\ref{eq:diff1}) 
cannot have  poles at  $Q^2 = -{\mathsf m}^2_{Ww}$.
From eqn. (\ref{eq:consistent}), we thus find
\begin{equation}
  \sum_{i=0}^N x_i 
    \left[ \prod_{\hat{i}=1}^i \left(1+ {Q^2\over {\mathsf m}_{i\,\hat{i}}^2}\right)\right]
  = {\cal A}(Q^2)
     \left[
       \prod_{w=1}^{N}
       \dfrac{Q^2 + {\mathsf m}^2_{W_{w}}}{{\mathsf m}^2_{W_{w}}}
     \right]~,
\label{eq:a}
\end{equation}
for some polynomial ${\cal A}(Q^2)$. The left hand side of
eqn. (\ref{eq:a}), however, is at most a polynomial
of degree $N$ in $Q^2$; therefore ${\cal A}(Q^2)$ must be a constant, 
which we denote ${\cal A}$ .

Applying eqn. (\ref{eq:a}), we may now evaluate 
$[G_{\rm NC}(Q^2)]_{WY}$ from eqns. (\ref{eq:gncwyii}) and
(\ref{eq:gncin}), and 
$[G_{\rm NC}(Q^2)]_{WW}-[G_{\rm CC}(Q^2)]_{WW}$
from eqns. (\ref{eq:usefulsum}) and (\ref{eq:consistencyi}), to find
\begin{equation}
  [G_{\rm NC}(Q^2)]_{WY} 
   = \dfrac{{\cal A}\,e^2 M_Z^2}{Q^2(Q^2+M_Z^2)}
     \left[
       \prod_{z=1}^{N}
       \dfrac{{\mathsf m}^2_{Z_{z}}}{Q^2 + {\mathsf m}^2_{Z_{z}}}
     \right]
     \left[
       \prod_{w=1}^{N}
       \dfrac{Q^2 + {\mathsf m}^2_{W_{w}}}{{\mathsf m}^2_{W_{w}}}
     \right],
     \label{eq:gncwya}
\end{equation}
and
\begin{eqnarray}
\lefteqn{
  [G_{\rm NC}(Q^2)]_{WW} 
 -[G_{\rm CC}(Q^2)]_{WW}  =
} \nonumber\\
 & & \dfrac{{\cal A}^2\,e^2 M_Z^2 M_W^2}{Q^2(Q^2+M_Z^2)(Q^2+M_W^2)}
     \left[
       \prod_{\hat{z}=1}^{N}
       \dfrac{{\mathsf m}^2_{Z_{\hat{z}}}}{Q^2 + {\mathsf m}^2_{Z_{\hat{z}}}}
     \right]
     \left[
       \prod_{w=1}^{N}
       \dfrac{Q^2 + {\mathsf m}^2_{W_{w}}}{{\mathsf m}^2_{W_{w}}}
     \right].
\label{eq:gncwwa}
\end{eqnarray}
Note that these expressions are specific to the case of an ideally delocalized fermion.
The normalization ${\cal A}$ can be determined by examining the residue of the
photon pole in either eqn. (\ref{eq:gncwya}) or (\ref{eq:gncwwa}). From
$[G_{\rm NC}(Q^2)]_{WY}$, for example, we have
\begin{equation}
[\xi_\gamma]_{WY} = {\cal A} e^2 = e^2\ \ \ \Rightarrow\ \ \ {\cal A}=1~,
\label{eq:photoncoupling}
\end{equation}
by charge universality, eqn. (\ref{eq:universality}).

We have previously shown \cite{SekharChivukula:2004mu,Chivukula:2004pk}
that $\alpha S$ may be computed by examining the residue of the
correlation function $[G_{\rm NC}(Q^2)]_{WY}$ at $Q^2=-M^2_Z$
\begin{equation}
  [\xi_Z]_{WY} = -e^2\left[1+ \dfrac{\alpha S}{4s^2 c^2}\right].
\end{equation}
Evaluating the residue from eqn. (\ref{eq:gncwya}), and 
expanding in inverse powers of the KK masses, 
we find
\begin{equation}
[\xi_Z]_{WY}=-e^2\,[1+M^2_Z(\Sigma_Z - \Sigma_W)]~,
\end{equation}
and therefore conclude
\begin{equation}
  \alpha S = 4s^2 c^2 M_Z^2 \left(\Sigma_Z - \Sigma_W\right)~.
\end{equation}
The parameter $\alpha T$ is independent of fermion delocalization, and
is computed in \cite{SekharChivukula:2004mu,Chivukula:2004pk} to be
\begin{equation}
  \alpha T = s^2 M_Z^2 \left(\Sigma_Z - \Sigma_W\right).
\end{equation}
In the continuum Higgsless models in warped and flat space discussed
in section 6 \cite{Csaki:2003zu,Foadi:2003xa}, we find \cite{Chivukula:unpublished} 
that $\alpha S$ and $\alpha T$ are 
small and slightly negative .

It is now straightforward to calculate the electroweak parameters of Barbieri et al. 
\cite{Barbieri:2004qk,Chivukula:2004af},
\begin{eqnarray}
  \hat{S} &=& \dfrac{1}{4s^2} \left(
    \alpha S + 4c^2 (\Delta\rho -\alpha T) + \dfrac{\alpha\delta}{c^2}
  \right), \\
  \hat{T} &=& \Delta\rho, \\
  W &=& \dfrac{\alpha\delta}{4s^2 c^2}, \\
  Y &=& \dfrac{c^2}{s^2} (\Delta\rho - \alpha T)~,
\end{eqnarray}
and for ideally delocalized fermions we obtain
\begin{equation}
  \hat{S} = \hat{T} = W = 0, \qquad
  Y = M_W^2 (\Sigma_W - \Sigma_Z).  
\end{equation}

\subsection{Normalization of Distribution of Ideally Delocalized Fermion}

The normalization ${\cal N}$ of the ideally delocalized fermion
in eqn. (\ref{eq:ideal}) can be determined from
the $W$-pole residue $[\xi_W]_{WW}$ as computed from 
eqn. (\ref{eq:gncwwa}). Expanding the residue in inverse powers of the KK masses, we find
\begin{equation}
[\xi_W]_{WW} = \left(\sum_i x_i g_i v^W_i\right)^2
= {e^2 \over \left(1-{M^2_W\over M^2_Z}\right)} \left[
1+M^2_W(\Sigma_Z - \Sigma_W)\right]~,
\end{equation}
and we recall that $[\xi_W]_{WW} = g^2_W$ from eqn. (\ref{eq:xivs}).
Taking the dot product of eqn. (\ref{eq:ideal}) with $v^W_i$, 
we see that the normalization ${\cal N}$ is equal to the
$W$ coupling to fermions
\begin{equation}
 {\cal N} = {\cal N} \sum_i v^W_i v^W_i = \sum_i x_i g_i v^W_i = g_W ~.
\end{equation}
Therefore
\begin{equation}
{\cal N}={e\over \sqrt{1-{M^2_W\over M^2_Z}}}\left[1+{M^2_W\over 2}
(\Sigma_Z - \Sigma_W)\right]~,
\end{equation}
and the profile of the ideally delocalized fermion is related to the $W$
wavefunction by
\begin{equation}
g_i x_i = g_W v^W_i~.
\end{equation}

\section{Two Examples in the Continuum}

We briefly discuss ideal fermion delocalization
in two continuum Higgsless models. In this paper, we will describe the
correspondence between the $SU(2)$ sectors of the deconstructed
and continuum models, and display the results for
the $W$ wavefunction and ideal fermion delocalization; details of the calculations 
and a discussion of the $U(1)$ sector of these models will be
presented in \cite{Chivukula:unpublished}.

\subsection{Warped Higgsless Models}

We first consider Higgsless models in Anti-deSitter space, as described
in \cite{Cacciapaglia:2004rb}. A deconstructed moose which yields this
model in conformally flat coordinates has the following parameters \cite{Chivukula:unpublished}
\begin{align}
f^2_i & = v^2\,(N+1)~,\label{eq:continuumf}\\
{1\over g^2_i} & = {1\over b g^2}\,\log\left[
{{i+1+(N-i)e^{-{b}}}\over{i+(N+1-i)e^{-b}}} \right]~,\ \ \ i=0,\ldots,N~.
\label{eq:warpedcoupling}
\end{align}
The continuum limit is taken by sending $N \to \infty$ while holding $g$, $v$,  and $b$ fixed.  
The conformally flat coordinates in the fifth dimension, $z$, 
corresponding to the deconstructed lattice points are
\begin{equation}
z_i = {R^\prime} \,\left[e^{-b}+{i\over N+1}(1-e^{-b})\right]^{1/2}~,
\label{eq:zi}
\end{equation}
and therefore in the continuum limit
\begin{equation}
2 z_i \Delta z_i = {{R^\prime}^2 \over N+1}\,(1-e^{-b}) \to 2 zdz~.
\label{eq:dz}
\end{equation}
For $i=0,1,\ldots,N+1$, we see that 
\begin{equation}
R\equiv R^\prime e^{-b/2} \le z \le R^\prime~,
\end{equation}
and hence, following \cite{Cacciapaglia:2004rb}, we identify $R\simeq 1/M_{pl}$ with the
position of the ``Planck'' brane and $R^\prime \simeq 1/{{\rm TeV}}$ with the position of
the ``TeV'' brane. We therefore expect 
\begin{equation}
b = 2\, \log{R^\prime \over R} \simeq {\cal O}(60)~.
\end{equation}
From eqns. (\ref{eq:warpedcoupling}), (\ref{eq:zi}), and (\ref{eq:dz}), we find that in
the continuum limit the coupling becomes position-dependent
\begin{align}
{1\over g^2_i} & = {1\over b g^2}\, \log\left[1+{{R^\prime}^2 \over N+1}
\left({{1-e^{-b}}\over z^2_i}\right)\right]~,\\
&\to {dz \over g_5^2(z)} = {2\,dz \over b g^2\, z}~.
\label{eq:gzcoupling}
\end{align}
Note that in eqn. (\ref{eq:gzcoupling}), we have interpreted the conformal factor
in the metric used to describe the model in \cite{Cacciapaglia:2004rb} as
a position-dependent five-dimensional gauge coupling $g^2_5 (z)$.

Requiring that the continuum limit of the nonlinear sigma model terms in
eqn. (\ref{lagrangian}) yield the $A_5$-dependent kinetic terms for a five-dimensional
Yang-Mills theory fixes the correspondence between the Goldstone bosons and
$A_5$, and implies that \cite{Arkani-Hamed:2001ca,Hill:2000mu}
\begin{equation}
\Delta z_i = {2 \over g_i f_i}~.
\end{equation}
Using
\begin{equation}
\sum_i \Delta z_i = R^\prime - R\approx R^\prime~,
\end{equation}
we then compute the relation
\begin{equation}
R^\prime = {4 \over \sqrt{b}\,gv}~.
\end{equation}
Solving the resulting continuum eigenvalue equation for the $W$
\cite{Cacciapaglia:2004rb,Chivukula:unpublished}, we find\footnote{Our
result for $M^2_W$ agrees with \cite{Cacciapaglia:2004rb} in the limit where
the coupling $g_R \to 0$.}
\begin{equation}
M^2_W \approx {g^2 v^2 \over 4} = {2 \over {R^\prime}^2 \log{R^\prime\over R}}~,
\end{equation}
and that near the Planck brane the $W$ wavefunction satisfies\footnote{Given the normalization of
the gauge kinetic-energy terms in eqn. (\ref{lagrangian}), it is the combination $g_i v^W_i$ that
has a well-defined continuum limit  \cite{Chivukula:unpublished}.}
\begin{equation}
g_i v^W_i \to g_5 (z) \chi^W(z) \simeq\, constant~,
\end{equation}
as expected from our discussion of the general properties of the $W$ eigenvector
in section \ref{sec:weigenvector}. 

Finally, we may compute the ideally delocalized
fermion wavefunction appropriate for this model. In the continuum
limit the fermion probabilities $x_i$ are related to the corresponding
continuum wavefunction,
$x_i \to |\psi(z)|^2 dz $, and hence near the Planck brane we find
\begin{align}
x_i & =  {g_W\,v^W_i \over g_i} = {g_W\,(g_i v^W_i) \over g^2_i}~, \nonumber \\
&  \to |\psi(z)|^2 dz  \propto  {2\,dz \over b\,z}~.
\end{align}
The ideally delocalized fermion wavefunction therefore corresponds 
near the Planck brane  to the ``flat'' wavefunction described
in \cite{Cacciapaglia:2004rb}. Since the flat and ideal
wavefunctions coincide near the Planck brane where the $W$ wavefunction is concentrated,  
the flat wavefunction results in $\hat{S}=\hat{T}=W=0$, 
and therefore $\alpha S=0$ \cite{Cacciapaglia:2004rb}, up to order $1/b^2$ \cite{Chivukula:unpublished}.

\subsection{Flat Higgsless Models}

Next we consider delocalized fermions in a flat background spacetime, as
discussed in \cite{Foadi:2004ps}. The deconstructed version of this model has
the same $f_i$ shown in eqn. (\ref{eq:continuumf}), but rather different gauge-couplings
\begin{equation}
{1\over g^2_i} =
\begin{cases}
{1\over g^2_0}~,\ \ \ i=0~,\\
{1 \over N\,\tilde{g}^2}~,\ \ \ i=1,2,\ldots,N~.\\
\end{cases}
\end{equation}
The continuum limit is taken holding $g_0$, $\tilde{g}$, and $v$ fixed, and the hierarchy of
masses between the $W$ and the KK modes is enforced by taking $g^2_0 \ll \tilde{g}^2$.
The fifth coordinates, $\tilde{x}$, corresponding to the deconstructed lattice points are
\begin{equation}
\tilde{x}_i = {i\,\pi R \over N+1}~,
\end{equation}
and, therefore, in the continuum limit
\begin{equation}
\Delta \tilde{x}_i = {\pi R \over N+1} \to d\tilde{x}~.
\end{equation}
For $i=0,1,\ldots,N+1$, we see that
\begin{equation}
0\le \tilde{x} \le \pi R~.
\end{equation}
The continuum limit of the gauge-couplings is then
\begin{equation}
{1\over g^2_i} \to \left({1\over \pi R\,\tilde{g}^2} + {1\over g^2_0} \delta(\tilde{x})\right)\,d\tilde{x}~,
\end{equation}
and the Lagrangian of eqn. (\ref{lagrangian}) yields a five-dimensional Yang-Mills
theory with coupling
\begin{equation}
g_5 = \tilde{g}\sqrt{\pi R}~,
\end{equation}
and  the necessary ``brane'' kinetic energy terms \cite{Foadi:2003xa}.

As in the previous example, requiring that the continuum limit of the nonlinear
sigma model terms in eqn. (\ref{lagrangian}) yield the $A_5$-dependent
kinetic energy terms for a five-dimensional Yang-Mills theory fixes the
correspondence between the Goldstone bosons and $A_5$, and implies
that \cite{Arkani-Hamed:2001ca,Hill:2000mu}
\begin{equation}
\Delta \tilde{x}_i = {2 \over g_i f_i}~.
\end{equation}
We therefore find the continuum limit
\begin{equation}
\sum_i \Delta {\tilde{x}}_i = \pi R \to  {2 \over \tilde{g} v}~.
\end{equation}
The $W$ wavefunction follows from the continuum eigenvalue equations
\cite{Foadi:2003xa,Chivukula:unpublished}, and we find
\begin{equation}
M^2_W \approx {g^2_0 v^2 \over 4}~,
\end{equation}
and
\begin{equation}
g_i v^W_i \to g_5 \chi^W(\tilde{x}) =
g_0\left[\left(1-{\tilde{x}\over \pi R}\right) - \left({g^2_0 \over 6\, \tilde{g}^2}\right)
\left(1-{\tilde{x} \over \pi R}\right)^3 +\ldots\right]~.
\end{equation}

The ideally delocalized fermion wavefunction in this model is then computed to
be
\begin{equation}
x_i \to |\psi(\tilde{x})|^2\, d\tilde{x} \propto  \left({1\over \pi R\, \tilde{g}^2} + {1\over g^2_0} \delta(\tilde{x})\right)g_5 \chi^W(\tilde{x})\,d\tilde{x}~,
\end{equation}
which, after normalization, yields
\begin{equation}
|\psi(\tilde{x})|^2\, d\tilde{x} = \left[
\left( 1-{g^2_0 \over 2\tilde{g}^2}\right)\delta(\tilde{x}) +
{g^2_0 \over\pi R\, \tilde{g}^2}\left( 1-{\tilde{x} \over \pi R}\right) + \ldots
\right]\,d\tilde{x}~.
\end{equation}
It may be verified that $\hat{S}=\hat{T}=W=0$ in this case; details 
will be given in \cite{Chivukula:unpublished}. 

Finally, we note that the 
slightly modified fermion distribution
\begin{equation}
|\phi(y)|^2\, dy = \left[
\left( 1-{B\,g^2_0 \over 2\tilde{g}^2}\right)\delta(\tilde{x}) +
{B\,g^2_0 \over\pi R\, \tilde{g}^2}\left( 1-{\tilde{x} \over \pi R}\right) + \ldots
\right]\,d\tilde{x}~,
\end{equation}
results \cite{Chivukula:unpublished} in a potentially nonzero value of $\alpha S$
\begin{equation}
{\alpha S \over 4 s^2 c^2} =
{(1-B) \over 6} \left[{g^2_0 \over \tilde{g}^2}
+ {g^2_{N+1} \over \tilde{g}^2}\right]~.
\end{equation}
In fact, this adjustable tree-level value of $\alpha S$ might
serve to cancel contributions arising from higher-order effects
\cite{Perelstein:2004sc}.

\section{Conclusions}

We have examined the properties of deconstructed Higgsless models
for the case of a fermion whose $SU(2)$ properties arise from delocalization over 
many sites of the deconstructed lattice.   We have derived expressions for the correlation functions 
and used these to establish a generalized consistency relation among them.  We have
discussed the form of the $W$ boson wavefunction and have shown that if the probability distribution
of the delocalized fermions is appropriately related to the $W$ wavefunction, the
precision electroweak parameters \cite{Barbieri:2004qk}  $\hat{S}$, $\hat{T}$, and $W$ are exactly
zero at tree-level.  

Our results may be applied to any Higgsless linear moose model with multiple 
$SU(2)$ groups, including a large class of extended electroweak gauge theories 
with only a few extra vector bosons. We have briefly discussed the ideal 
fermion  delocalization in two continuum Higgsless models,  one in Anti-deSitter 
space  and one in flat space. The details of  deconstructing the continuum flat 
and Anti-deSitter space Higgsless models, and of computing the ideal fermion 
delocalization distributions  and a discussion of the $U(1)$ sectors of these 
models, will be presented in \cite{Chivukula:unpublished}.

\acknowledgments

R.S.C. and E.H.S. are supported in part by the US National Science Foundation under
award PHY-0354226, and gratefully acknowledge the hospitality of the Tohoku University
Theory group. M.K. is supported by a MEXT Grant-in-Aid for Scientific Research
No. 14046201.
M.T.'s work is supported in part by the JSPS Grant-in-Aid for Scientific Research No.16540226. H.J.H. is supported by the US Department of Energy grant
DE-FG03-93ER40757.



\end{document}